\documentstyle[preprint,aps]{revtex}

\begin{document}
\draft
\title{Negative differential conductivity and population inversion in the
double-dot system connected to three terminals}
\author{Lev G. Mourokh}
\address{Department of Physics and Engineering Physics, \\
Stevens Institute of Technology, Hoboken, NJ 07030 }
\author{Anatoly Yu. Smirnov}
\address{D-Wave System, Inc., 320-1985 W. Broadway, Vancouver, \\
British Columbia, Canada V6J 4Y3}
\date{\today }
\maketitle

\begin{abstract}
{We examine transport and microwave properties of two coupled quantum dots taken in parallel connection to the common left lead and connected to separate leads at their right side. In addition, the area between the left lead and the double-dot structure is threaded by Aharonov-Bohm magnetic flux. We determine the energies and populations of double-dot levels on the microscopic basis taking into account the interdot Coulomb interaction and show that at large lead-to-lead bias the population inversion can be achieved. For the case of strong Coulomb repulsion, this inversion leads to level crossing accompanied by the region of negative differential conductivity in the current-voltage characteristics, whereas for weaker Coulomb repulsion, the resonant microwave absorption becomes negative at high lead-to-lead voltage.}
\end{abstract}


\narrowtext

Recent achievements in nanotechnology have led to a new generation of
semiconductor devices based on quantum dots that can be viewed as artificial
atoms and multiple quantum dot systems playing the role of artificial
molecules \cite{1}. In these devices the quantum properties of carriers are explored and the Coulomb interaction is of crucial importance in such small devices. Initially, in most studies, the dots were taken to be
connected in series, but, recently, double-dot structures in which the two
constituent dots are placed in a parallel arrangement between leads have
been also analyzed theoretically \cite{LS,SHM} as well as experimentally 
\cite{Exp}. Such a two-dimensional system can be threaded by Aharonov-Bohm
(AB) magnetic flux to examine interference effects. In particular, it was
shown \cite{SHM} that for a symmetric double-dot system having molecular
bonding and antibonding states, one of these states can be disconnected from
the leads at appropriate values of the AB flux, $\Phi$. These values are $%
\Phi /\Phi_0 = 2n$ for the antibonding state and $\Phi /\Phi_0 = 2n+1$ for
the bonding state, respectively, where $\Phi_0 = hc/e$ is the flux quantum
and $n = 0,1,2,...$. The electron spin entangler based on the {\it triple}%
-dot system with separate leads connected to each of the dots was proposed
in Ref. \cite{SL}.

In this work we examine a system combining the properties of structures
analyzed in Refs. \cite{SHM} and \cite{SL}, i.e. the double-dot connected to
the same lead in their common left side in parallel arrangement, while in
their right side the two dot are connected to the separate leads (Figure 1).
In addition, the left side of structure is threaded by AB flux which can be
used to control the connection of the double-dot system to the left lead. We determine the double-dot level populations and show that, in contrast to the case of symmetric double-dot coupling to the same leads both at the left and at the right sides of the structure \cite{SHM}, they exhibit dependence on the AB magnetic flux. Accordingly, the population of the antibonding state can be larger at high lead-to-lead bias than the population of the bonding state, and either the antibonding state becomes the ground state of the system or the population inversion is achieved. We also analyze the lead-to-lead current and show that, in the case of strong interdot Coulomb coupling, level crossing is accompanied by the region with negative differential conductivity on the current-voltage characteristics.

The second-quantized Hamiltonian of the double-dot electrons including the
interdot Coulomb interaction is given by 
\begin{equation}
H_{0}=E_{0}(a_{1}^{+}a_{1}+a_{2}^{+}a_{2})-\Delta
(a_{1}^{+}a_{2}+a_{2}^{+}a_{1}) + Ua_{1}^{+}a_{2}^{+}a_{2}a_{1},
\end{equation}
where $a_{i}^{+},a_{i}$ are the creation/annihilation operators for
electrons in $i$-th dot $(i=1,2)$, $-\Delta $ is a tunneling constant
between the dots, $U=e^{2}/2d\varepsilon (\varepsilon $ is the dielectric
constant). The Hamiltonian of the leads has the form 
\begin{equation}
H_{Leads}=\sum_{k}E_{Lk}c_{Lk}^{+}c_{Lk} + \sum_{k}E_{R1k}c_{R1k}^{+}c_{R1k}
+ \sum_{k}E_{R2k}c_{R2k}^{+}c_{R2k},
\end{equation}
where $c_{\alpha k}^{+}(t),c_{\alpha k}(t)$ are creation/annihilation
operators of electrons with momentum ${\bf k}$ in the $\alpha$-lead ($%
\alpha =L,R_1,R_2$). The effect of AB flux on quantum transport can be taken
into account using the Peierls gauge phase factors $exp(\pm i\phi )$ in a
transfer matrix description of tunneling between the left lead and dots,
with $\phi = Bld/4\Phi _{0}$ is the phase experienced by an electron during
the lead-double-dot tunneling process. With this, the Hamiltonian for
tunneling between dots and leads is written as 
\begin{equation}
H_{tun} = \sum_{k}L_{k}c_{Lk}^{+}(a_{1}e^{i\phi }+a_{2}e^{-i\phi
})+\sum_{k}R_{1k}c_{R1k}^{+}a_{1}+\sum_{k}R_{2k}c_{R2k}^{+}a_{2} + h.c.
\end{equation}

Employing the procedures of Ref.\cite{SHM}, we derive and solve the
equations of motion for Green's functions of the double-dot electrons,
obtaining a self-consistent set of equations for the level populations, 
\begin{equation}
N_{A,B}={\frac{f_{L}(E_{A,B})(1\pm\cos 2\phi )}{2\pm\cos 2\phi}} +{\frac{%
f_{R1}(E_{A,B})+ f_{R2}(E_{A,B})}{2(2\pm\cos 2\phi) }},
\end{equation}
and the level energies 
\begin{equation}
E_{A,B}=E_0 \pm\Delta + UN_{B,A},
\end{equation}
where $f_{\alpha }(E_{A,B})$ are the distribution functions of electrons in
the left and right leads taken at the energies of the double-dot levels as 
\begin{equation}
f_{\alpha }(E_{A,B})=\left[ \exp \left( {\frac{E_{A,B}-\mu -(-1)^{\epsilon
_{\alpha }}eV/2}{T}}\right) +1\right] ^{-1},
\end{equation}
where $\epsilon _{L}=0,\epsilon _{R_1,R_2}=1,V$ is the lead-to-lead bias
voltage (assuming the same bias to be applied to both right leads) and $\mu $
is the equilibrium chemical potential common to all three leads. It should
be noted that the magnetic field induced phase is not cancelled in Eq. (4)
as it was in the case of symmetrical connections to the single left and
right leads \cite{SHM}, where the populations were given by $N_{A,B}=(f_L(E_{A,B})+ f_L(E_{A,B}))/2$. In the latter case the limiting high-bias values for the populations of both levels were 1/2 with the population of the antibonding state approaching this value from above and the population of the bonding state approaching this value from below. For the case examined in the present paper, the populations are magnetic flux dependent and their limiting values are determined by the AB phase.  Moreover, the population of the antibonding state
can be larger than the population of the bonding state at appropriate values
of the applied magnetic field and lead-to-lead bias, which can lead to the population inversion or,
for strong Coulomb repulsion, to the situation when the antibonding state
becomes the ground state of the system. It should be emphasized that the population inversion can be achieved only for the strongly nonequilibrium situation. For example, at zero bias the distribution functions are the same for all leads, the AB phase dependence is canceled, and the level populations are determined by this distribution taken at the energy of the corresponding level. Consequently, the population of the antibonding state is always less than the population of the bonding state.

Eqs. (4) and (5) were solved self-consistently with the voltage dependencies
of level populations and energies shown in Figure 2(a) and Figure 2(b),
respectively, for $\phi = \pi /8$, low temperature $T=0.2\Delta$, large
Coulomb energy $U=8\Delta$, and separation between the equilibrium chemical
potential of the leads, $\mu$, and the energy of the single dot ground
state, $E_0$, chosen as $7\Delta$. In this case, in equilibrium the bonding
state is below the Fermi level and the antibonding state is above the Fermi
level with initial populations to be $N_B=1$ and $N_A=0$. With voltage
increasing, the chemical potential of the left lead passes through the
antibonding state (modulo thermal broadening) resulting in its population
(Figure 2(a)). Accordingly, the energy of the bonding state increases
(Figure 2(b)) and with further voltage increasing the chemical potential of
the right leads passes through the energy of this state resulting in its
depopulation and corresponding decrease of the antibonding state energy. The
Coulomb energy is chosen to be sufficiently large, so that the energy of the
antibonding state becomes less than the energy of the bonding state and,
moreover, less than the chemical potential of the right lead. Consequently,
this level becomes nonconductive and its population increases up to one. With further voltage increasing the antibonding state remains the ground state of the system and becomes conductive again.

These population changes manifest themselves in the current-voltage
characteristics of the double-dot system. For symmetric coupling to the
leads and low temperature the current through the structure is given by 
\begin{eqnarray}
I &=&{\frac{e\Gamma }{\hbar }}\left[ \frac{1+\cos 2\phi }{2+\cos 2\phi }%
\left( f_{L}(E_{A})-\frac{f_{R1}(E_{A})+f_{R2}(E_{A})}{2}\right) +\right. \nonumber \\ 
&&\left. \frac{1-\cos 2\phi }{2-\cos 2\phi }\left( f_{L}(E_{B})-\frac{%
f_{R1}(E_{B})+f_{R2}(E_{B})}{2}\right) \right] 
\end{eqnarray}
where $\Gamma $ is the lead-dot coupling constant. The current-voltage characteristics for the above listed set of parameters is presented in Figure 3. It is evident from this Figure that there sharp drop in the current at voltage when the antibonding state become nonconductive and, correspondingly, there is the region with negative differential conductivity.

The case of weaker Coulomb repulsion ($U=2\Delta$) is also of interest. (It should be noted that the ratio of the tunnel coupling and the Coulomb energy can be varied in wide range by the change of the interdot distance.) The level populations and energies for this situation are presented in Figure 4 (a) and (b), respectively, as functions of the applied bias voltage. It is evident from these figures that the bonding state remains the ground state of the system but there is the population inversion at high voltage.

To further examine the properties of the double-dot structure, we analyze the resonant microwave absorption. The energy absorbed by the system is given by \cite{SHM}
\begin{equation}
P=(E_A-E_B)(N_B-N_A),
\end{equation}
and we can expect the amplification of the incident microwave field when this parameter becomes negative. The voltage dependence of the absorbed energy is shown in Figure 5 (a) for $U=8\Delta$ and (b) for $U=2\Delta$. One can see that for the former case there is small region when $P<0$ corresponding to the conditions of negative differential conductivity. However, for the weaker Coulomb energy (Figure 5(b)) the system can amplify the external field for the voltages higher than some threshold value. This amplification can be controlled by the magnetic field as can be seen in Figure 6 where the absorbed energy is plotted as a function of the AB phase $\phi$.

In conclusion, we have examined the transport and microwave properties of the double-dot structure connected in parallel to the same terminal from its left side and having the two dots connected to the separate terminals from their right side. In addition, the left side of the structure is threaded by Aharonov-Bohm magnetic flux. We have shown that both level populations and electric current through the structure are functions of AB flux and at appropriate choice of parameters both the negative differential conductivity and the population inversion can be achieved with possible amplification of external microwave field.

\newpage
\begin{center}
Figure Captions
\end{center}

Figure 1. Schematic of the double-dot system with connections to three terminals.

Figure 2. (a) Level populations and (b) energies as functions of the applied voltage bias for the case of strong Coulomb repulsion.

Figure 3. Current-voltage characteristics of the structure.

Figure 4. (a) Level populations and (b) energies as functions of the applied voltage bias for the case of weak Coulomb repulsion.

Figure 5. Absorbed resonant microwave energy as function of the applied voltage bias for (a) strong Coulomb repulsion and (b) weak Coulomb repulsion.

Figure 6. Absorbed resonant microwave energy as function of the Aharonov-Bohm phase.

\end{document}